\documentstyle[12pt]{article}

\advance \textheight by \topskip
\newcommand{\be}{\begin{equation}}
\newcommand{\ee}{\end{equation}}
\newcommand{\bea}{\begin{eqnarray}}
\newcommand{\eea}{\end{eqnarray}}
\newcommand{\nn}{\nonumber}
\newcommand{\e}{\epsilon}
\newcommand{\s}{\sigma}
\newcommand{\Si}{\Sigma}
\newcommand{\ov}{\overline}
\newcommand{\un}{\underline}
\newcommand{\st}{space-time }

\title{
{\bf Symplectic vs pseudo-Euclidean}\\
{\bf space-time with extra dimensions\thanks{Report presented at the
XI-th Int.\
School ``Particles and Cosmology'', Baksan Valley, 18-24 April,
2001.}}
}
\author{Yu.\ F.\ Pirogov\\[0.5ex]
{\it IHEP, Protvino, Moscow Region RU-142284, Russia}}
\date{}
\begin{document}
\maketitle

\abstract{
\noindent
It is conjectured that the symplectic structure of \st is superior to
the metric one. Instead of the commonly adopted pseudo-orthogonal
groups $SO(1,d-1)$, $d\ge 4$, the complex symplectic ones $Sp(2l,C)$, 
$l\ge 1$ are proposed as the local structure groups of the extended
space-time. A~discrete series of the metric space-times of the
particular dimensionalities $d=4l^2$ and
signatures, with $l(2l-1)$ time and  $l(2l+1)$ spatial
directions, defined over the set of the Hermitian second-rank
spin-tensors is proposed as an alternative to the pseudo-Euclidean
space-times with extra dimensions. The one-dimensional time-like
direction remaining invariant under fixed boosts makes it possible
the non-relativistic causality description despite the presence of
extra times.
}

\section{Space-time: symplectic vs pseudo-Euclidean}

The \st we live in is generally adopted to be (locally) the 
Minkowski one. Nevertheless, the spinor calculus in this \st
heavily relies on the isomorphism of the noncompact groups
$SO(1,3)\simeq SL(2,C)/Z_2$, as well as that $SO(3)\simeq SU(2)/Z_2$
for their maximal compact subgroups. In fact, the whole relativistic
field theory in four \st dimensions can equivalently be formulated
in the framework of representations of the complex unimodular
group $SL(2,C)$ alone (and in a sense it is  even
preferable~\cite{pen}).
For this reason, the \st structure group
with spinors as defining representations, i.e.\ the complex symplectic
group $Sp(2,C)\simeq SL(2,C) $, is  conceptually more appropriate
than the pseudo-orthogonal one $SO(1,3)$ with vectors as defining
representation.  In the former approach, to a
\st point there corresponds  a Hermitian spin-tensor of the second
rank rather then a four-vector.

Then in searching for the space-times with  extra dimensions it is
natural to look for the extensions in the symplectic framework with
the structure group $Sp(2l,C)$, $l>1$ instead of $SO(1,d-1)$, $d>4$.
The symplectic series of the groups is peculiar quantum-mechanically
for it retains the bilinear spinor product at any $l>1$. 
Two alternative ways of the space-time extension can
be pictured  schematically as follows:
\bea
SO(1,3)&\simeq& Sp(2,C) \nn \\
\downarrow\ \ \ \ &&\ \ \ \ \ \downarrow\nn \\
SO(1,d-1)&\not\simeq& Sp(2l,C)\,. \nn
\eea
The first, commonly adopted way of extension, corresponds to the
pseudo-orthogonal structure groups while the second one relies on the
complex symplectic groups.
The scheme shows that the isomorphism of the two types of  groups,
valid at $d=4$ and  $l=1$, is no longer fulfilled at $d>4$ and  $l>1$.
In the first way of extension the local metric properties of
the space-times, i.e.\ their dimensionalities and signatures, are put
in from the very beginning. In the second way, these properties are
not to be considered as the primary ones but, instead, they should
emerge as a manifestation of the inherent symplectic structure.
In what follows, we develop the  symplectic framework in general  
and elaborate somewhat the ordinary and next-to-ordinary space-time
cases with $l=1,2$, respectively.\footnote{For more detail
see~\cite{pir}.}

\section{Symplectic \st}

Let $\psi_A$ and  $\bar\psi{}^{\bar A}\equiv (\psi_A)^*$, as
well as their respective duals $\psi^A$  and  $\bar\psi{}_{\bar
A}\equiv (\psi^A)^*$, $A$, $\bar A =1,\dots, 2l$ are the 
spinor representaions of $Sp(2l,C)$. There exist 
the invariant spin-tensors $\e_{AB}=-\e_{BA}$ and $\e^{AB}=-\e^{BA}$
such that $\e_{AC}\e^{CB}=\delta_A{}^B$, with $\delta_A{}^B$ being the
Kroneker symbol
(and similarly for $\e_{\bar A\bar B}\equiv(\e^{BA})^*$ and  $\e^{\bar
A\bar B}\equiv(\e_{BA})^*$). Owing to these  tensors the
spinor indices of the upper and lower positions  are pairwise
equivalent ($\psi_A\equiv\e_{AB}
\psi^B$ and $\bar \psi_{\bar A}\equiv \e_{\bar
A\bar B}\bar \psi^{\bar B}$), so that there are left just two
inequivalent spinor
representaions (generically, $\psi$ and $\bar\psi$). 
They are the spinors of the first and the second kind, respectively.

Let us put in correspondence to an event point $P$ a second rank
$2l\times 2l$ spin-tensor $X_A{}^{\bar B}(P)$, which is Hermitian,
i.e.,  fulfil the restriction 
$$X_A{}^{\bar B}= (X_B{}^{\bar A})^*\equiv\bar X^{\bar B}{}_A\,,$$
or in other terms $X^{A\bar B}=
(X_{B\bar A})^*$.  Now, one can define the quadratic scalar product
$$\label{eq:X2}
(X,X)\equiv\mbox{tr\,}X\bar X= X_A{}^{\bar B}\bar
X_{\bar B}{}^A= 
-X_{A\bar B}(X_{B\bar A})^*\,,
$$
with $(X,X)$ being  real though not sign definite. 
Under arbitrary $S\in Sp(2l,C)$ one has in short notations:
\bea
X&\to &SXS^\dagger\,,\nn\\
\bar X&\to &S^{\dagger -1}\bar X S^{-1}\,,\nn
\eea
so that $X\bar X \to S X\bar X S^{-1}$ and hence $(X,X)$ is invariant.
At $l>1$, the quadratic invariant above is just the
lowest order one in a series of independent invariants
$\mbox{tr\,}(X\bar X)^k$, $k=1,\dots,l$, the highest order one with
$k=l$ being equivalent to $\mbox{det}X$.

\vspace{1ex}
\noindent
{\bf Definition:}
{\it
the Hermitian spin-tensor set  $\{X\}$ equipped with the structure
group $Sp(2l,C)$ and the
interval between points $X_1$ and $X_2$ equal to
$(X_1-X_2,X_1-X_2)$ constitutes the flat symplectic space-time.
}

\vspace{1ex}
The noncompact transformations from the
$Sp(2l,C)$ are counterparts of the Lorentz boosts in the ordinary
space-time, while transformations
from  the compact subgroup $Sp(2l)=Sp(2l,C)\cap SU(2l)$ correspond to
rotations. With account for translations $X_A{}^{\bar B}\to
X_A{}^{\bar B}+\Xi_A{}^{\bar B}$, where  $\Xi_A{}^{\bar B}$ is an
arbitrary constant Hermitian spin-tensor,  the whole theory in the
flat symplectic space-time  should be covariant relative to the
inhomogeneous symplectic group. 

With restriction by the  the maximal compact subgroup $Sp(2l)$, the
indices of the first and the second kinds in  the same position are
indistinguishable  relative to their 
transformation properties ($\psi_A\sim \bar\psi_{\bar A}$, $\psi^A\sim
\bar\psi^{\bar A}$). One
can temporarily label $X_{A\bar B}$ in this case
as $X_{XY}$, where $X,Y,\dots=1,\dots,2l$ generically mean spinor
indices irrespective of their kind. Hence,  one can reduce the
tensor $X_{XY}$ into two  irreducible parts, symmetric and
antisymmetric ones: $X_{XY}=\sum_{\pm}(X_{\pm})_{ XY}$, where
$(X_{\pm})_{XY}=\pm (X_{\pm})_{YX}$ have the dimensionalities
$d_\pm=l(2l\pm 1)$, respectively. One gets  the following
decomposition
$$
(X, X)=\sum_{\pm} (\mp 1)(X_{\pm})_{XY}[(X_{\pm})_{XY}]^*\,. 
$$ 
At $l>1$, one can further
reduce the antisymmetric spin-tensor $X_-$ into the trace  relative to
$\e$
and a traceless part. Therefore the whole
extended space-time can be decomposed relative to the rotational
subgroup  into three irreducible subspaces of the dimensionalities
$1$, $(l-1)(2l+1)$ and $l(2l+1)$, respectively. The first two
subspaces correspond to time-like  directions, the one-dimensional
rotationally invariant and non-invariant ones, while the
third subspace corresponds to the spatial extra dimensions. In the
ordinary space-time
which corresponds to $l=1$ the second subspace is empty.

The particular decomposition of $X$ into two parts
$X_{\pm}$ is noncovariant with respect to the whole $Sp(2l,C)$ and
depends on the boosts. Nevertheless, the decomposition being valid at
any boost, the number of the positive and negative
components in $(X,X)$ is invariant under the whole
$Sp(2l,C)$. In other words,  the metric tensor of the flat symplectic
\st 
$$
\eta_d=(\,\underbrace{+1,\dots}_{d-}\,;\underbrace{-
1, \dots}_{d_+}\,) 
$$
is invariant. Hence, at $l>1$ the structure
group $Sp(2l,C)$ of the  $2l$-th rank  and the  $2l(2l+1)$-th order, 
acting on the Hermitian second-rank spin-tensors with  $d=4l^2$
components, is a subgroup of the embedding
pseudo-orthogonal group $SO(d_-,d_+)$, of the rank $2l^2$ and the
order $2l^2(4l^2-1)$,  acting on the  pseudo-Euclidean
space of the  dimensionality $d=4l^2$. What distinguishes $Sp(2l,C)$
from $SO(d_-,d_+)$, is the total set of independent  invariants
$\mbox{tr}(X\bar X)^k$, $k=1,\dots,l$. The isomorphism between
the  groups is valid only at $l=1$, i.e., for the 
ordinary space-time $d=4$ where there is just one independent
invariant $1/2\, \mbox{tr}X\bar X=\mbox{det}X$.

In the symplectic  approach we consider, neither the discrete set of
dimensionalities,
$d=4l^2$, of the extended space-time, nor its  signature, nor the
existence of the rotationally invariant one-dimensional time subspace
are postulated from the beginning. Rather, these properties are the
attributes of the underlying symplectic structure. In particular,  the
latter one  seems to provide at the fundamental level the simple
rationale for the four-dimensionality of the ordinary space-time, as
well as for its signature ($+---$). Namely, the last properties just
reflect the existence of one antisymmetric and three symmetric
second-rank Hermitian spin-tensors at $l=1$.
The set of such tensors, in its turn,  is the lowest admissible
Hermitian space to accommodate the symplectic structure. On the other
hand, right the  one-dimensionality of time is what allows the
events to be ordered at any fixed boosts and hence insures the causal
description. Therefore the latter one may ultimately be attributed to
the underlying symplectic structure, too. At $l>1$, because of the
one-dimensional time being mixed via boosts with the extra times,
the causality is  expected to be violated at large boosts.

\section{{\boldmath C, P, T} symmetries}

Let us  charge double  the spinor space, i.e., for
each  $\psi_A$ and $(\psi_A)^\dagger\equiv\bar\psi^{\bar A}$ introduce
two copies $\psi_A^{\pm}$ and  
$(\psi^\pm_A)^\dagger \equiv (\bar\psi^\mp)^{\bar A}$, with
$\pm$ being the ``charge'' sign. We use here a dagger sign
for complex conjugation to show that the Grassmann fields should
undergo the change of the order in their products.  
In analogy to the ordinary
case of $SL(2,C)$~\cite{streater}, one can define the following
discrete symmetries:
\bea\label{eq:CPT}
C&:&\psi^{\pm}_A\to \psi^{\mp}_A\,,\nn\\
P&:&\psi^{\pm}_A\to ({\psi^{\mp}_A})^\dagger\equiv
(\ov {\psi}{}^{\pm})^{\bar A}\,,\nn\\
T&:&\psi^{\pm}_A\to ({\psi^{\pm}_A})^\dagger\equiv
(\ov {\psi}{}^{\mp})^{\bar A}\,,\nn
\eea
and hence $CPT: \psi^{\pm}_A\to \psi^{\pm}_A$ (all up to
the phase factors). Under validity of the $CPT$ invariance, only two
of the discrete operations are independent ones.
Without charge doubling, just one independent combination
$CP\equiv T : \psi_A\to\bar\psi{}^{\bar A}$ survives. 

Now, let us introduce the Hermitian spin-tensor current
$J =J^\dagger$ as follows
$$\label{eq:J}
J_{A}{}^{\bar B}\equiv\sum_{\pm} (\pm 1)
\psi^{\pm}_{A}({\psi}{}^\pm_{B})^\dagger
=\sum_{\pm} (\pm 1)
\psi^{\pm}_{A}(\overline{\psi}{}^\mp){}^{\bar B}\,.
$$
($\psi$'s are the Grassmann fields). Under the discrete operations
above  the current
$J_A{}^{\bar B}$ transforms as follows 
\bea\label{eq:J_AB}
C&:&J_A{}^{\bar B}\to -J_A{}^{\bar B}\,,\nn\\
P&:&J_A{}^{\bar B}\to - J_B{}^{\bar A}\,,\nn\\
T&:&J_A{}^{\bar B}\to \phantom{-} J_B{}^{\bar A}\,.\nn
\eea
Fixing  boosts and decomposing current $J_{A\bar B}$ into  the
symmetric and antisymmetric parts,
$J_{XY}=\sum_{\pm}(J_{\pm}){}_{XY}$, one gets
\bea
C&:&(J_{\pm})_{XY}\to -(J_{\pm})_{XY}\,,\nn\\
P&:&(J_{\pm})_{XY}\to \mp (J_{\pm}){}^{XY}\,,\nn\\
T&:&(J_{\pm})_{XY}\to \pm (J_{\pm}){}^{XY}\,.\nn
\eea
This is in complete agreement with the signature association for the
symmetric (antisymmetric) part of the Hermitian
spin-tensor $X$ as the extended spatial (time) components.

\section{{\boldmath l = 1} space-time}

The noncompact group $Sp(2l,C)$ has $2l(2l+1)$ generators
$M_{AB}=(L_{AB}, K_{AB})$, $A,B=1,\dots,2l$, with
$M_{AB}=M_{BA}$. The  generators $L_{AB}$ are Hermitian and 
correspond to the extended rotations, whereas those $K_{AB}$ are
anti-Hermitian and correspond to the extended boosts. In the
space of the first-kind spinors $\psi_A$ these generators can be
represented as
$(\s_{AB}, i\s_{AB})$ with 
$(\s_{AB})_{CD}= 1/2 (\e_{AC}\e_{BD}+\e_{AD}\e_{BC})$, so that
$\s_{AB}=\sigma_{BA}$ and 
$(\s_{AB})_{CD}=(\s_{AB})_{DC}$, $(\s_{AB})_{C}{}^C=0$.
Similar expressions hold true in the space of the second-kind spinors
$\bar \psi_{\bar A}$. In these terms, a canonical formalism can be
developed at arbitrary $l\ge 1$.

However, in the simplest  case  $l=1$, corresponding to the ordinary
four-dimensional space-time, there exists the isomorphism
$SO(3,C)\simeq Sp(2,C)/Z_2$. Due to this property, the structure of
$Sp(2,C)$ can be brought to the form more familiar physically, though
equivalent mathematically. We use here the complex group 
$SO(3,C)$ instead of the real one $SO(1,3)$ to clarify the close
similarity with the subsequent case $l=2$ where there is no real
structure group. Because of the complexity
of $SO(3,C)$ one should distinguish vectors and their complex
conjugate, the latter ones being omitted for simplicity in
what follows. The same remains true for the $SO(5,C)$ case 
corresponding to $l=2$.

Let us introduce for the $SO(3,C)$
group the double set of the Pauli matrices, $(\s_{i})_{A}{}^{\bar B}$
and $(\overline\s_{i})_{\bar A}{}^{B}$,
$i=1,2,3$. They should satisfy the anticommutation relations:
$\s_i\overline \s_j+\s_j\overline \s_i=2 \delta_{ij}\s_0$ and
$\ov\s_i \s_j+\ov\s_j \s_i=2 \delta_{ij}\ov\s_0$, where
$(\s_{0})_{A}{}^B\equiv \delta_A{}^B$, 
$(\ov\s_{0})_{\bar A}{}^{\bar B}\equiv \delta_{\bar A}{}^{\bar B}$ are
the Kroneker symbols and $\delta_{ij}$ is the metric tensor of
$SO(3,C)$. Among these matrices,  $\s_0$ and $\ov\s_0$  are the
only independent ones which can be chosen antisymmetric:
$(\s_0)_{AB}\equiv
\e_{AB}$ and  $(\ov\s_0)_{\bar A\bar B}\equiv \e_{\bar A\bar B}$.
On the other hand, with respect to the maximal compact subgroup
$SO(3)$, all the matrices $\s_i$, $\ov\s_i$ can be chosen
both Hermitian and symmetric as $(\s_i)_X{}^{Y}=[(\s_i)_Y{}^{X}]^*$
and $(\s_i)_{X Y}=(\s_i)_{YX}$ (and the same for $\ov\s_i$). 
The  matrices $\s_{ij}\equiv -i/2\,(\s_i\ov\s_j-\s_j\ov\s_i)$
satisfying $\s_{ij}=-\s_{ji}$ and  $(\s_{ij})_{AB}=(\s_{ij})_{BA}$
(and similarly for $(\ov\s_{ij})_{\bar A\bar B}\equiv
i/2\,(\ov\s_i\s_j-\ov\s_j\s_i)_{\bar A\bar B}$) are
not linearly independent from $\s_i$. They can be brought to the form
$(\s_{ij})_{XY}=\e_{ijk}\,(\s_k)_{XY}$, with $\e_{ijk}$ being the
Levi-Civita $SO(3,C)$ symbol. 

The matrices ($\s_{ij}, i\s_{ij}$) can be identified as the generators
$M_{ij}=(L_{ij}, K_{ij})$ of the noncompact $SO(3,C)$ group in the
space of the first-kind spinors.  Respectively, in the space of the
second-kind spinors
they are ($-\ov\s_{ij}, i\ov\s_{ij}$). The generators $L_{ij}$
of the maximal compact subgroup $SO(3)\simeq Sp(2)/Z_2$ correspond to
rotations, while those $K_{ij}$ of the noncompact
transformations describe Lorentz boosts. 
Relative to the maximal compact subgroup $SO(3)$ one has $\bar
\s_0=\s_0$, $\bar \s_i=\s_i$ and $\bar \s_{ij}=-\s_{ij}$. When
restricted by $SO(3)$, the Hermitian second-rank
spin-tensor may be decomposed in the complete set of the Hermitian
matrices ($\s_0, \s_{ij}$) with the real
coefficients: $X=1/\sqrt 2\,
(x_0\s_0+1/2\,x_{ij}\s_{ij})$, so that 
$(X,X)=x_0^2-1/2\,x_{ij}^2$.
Identifying  $x_{ij}\equiv \e_{ijk}x_k$ one gets as
usually $(X,X)=x_0^2-x_{i}^2$.
Both the time and spatial representations being irreducible under
$SO(3)$, there takes place the usual decomposition 
$\underline 4=\underline 1\oplus \underline 3$ 
relative to the embedding $SO(3,C)\supset SO(3)$.

\section{{\boldmath l = 2} space-time}

This case corresponds to the next-to-ordinary \st symplectic
extension. There takes place the  isomorphism
$SO(5,C)\simeq Sp(4,C)/Z_2$. Cases $l=1, 2$ are
the only ones when the  structure of the symplectic group  gets
simplified in terms of the complex orthogonal groups. 
Relative to the
maximal compact subgroup $SO(5)$, the double set of Clifford  matrices
$(\Si_I)_A{}^{\bar B}$ and $(\ov \Si_I)_{\bar A}{}^{B}$, $I=1,\dots,5$
can  be chosen as Hermitian 
$(\Si_I)_X{}^{Y}= [(\Si_I)_Y{}^{X}]^*$ and antisymmetric
$(\Si_I)_{XY}=-(\Si_I)_{YX}$ (and similarly for
$\ov \Si_I$), like $(\Si_0)_{AB}=\e_{AB}$ and $(\ov\Si_0)_{\bar A\bar
B}=\e_{\bar A \bar B}$. One can also require that $(\Si_I)_X{}^X=0$
and $(\ov\Si_I)_X{}^X=0$. Thus under restriction by
$SO(5)$, six matrices $\Si_0$,  $\Si_I$  provide the complete
independent set for the antisymmetric matrices in the
four-dimensional spinor space. After introducing 
matrices $\Si_{IJ}=-i/2 (\Si_I\ov\Si_J-\Si_J\ov \Si_I)$, so that
$\Si_{IJ}=-\Si_{JI}$, one gets the  symmetry condition for them:
$(\Si_{IJ})_{AB}= (\Si_{IJ})_{B A}$ (and similarly for
$(\ov\Si_{IJ})_{\bar A\bar B}=i/2 (\ov\Si_I\Si_J-\ov\Si_J
\Si_I))_{\bar A\bar B}$. Therefore ten matrices
$\Si_{IJ}$  make up the complete set for the symmetric
matrices in the spinor space. The  matrices ($\Si_{IJ}$, $i\Si_{IJ}$) 
represent the $SO(5,C)$
generators $M_{IJ}=(L_{IJ},K_{IJ})$ in the space of spinors of
the first  kind, whereas  matrices ($-\ov\Si_{IJ}$, $i\ov\Si_{IJ}$)
do the job in the space of the second kind spinors.  

With respect to 
$SO(5)$ the Hermitian second-rank spin-tensor $X$ may be decomposed in
the complete set of matrices $\Si_0$, $\Si_I$ and $\Si_{IJ}$ with the
real coefficients: $X=1/2\,(x_0\Si_0+x_I\Si_I+ 1/2\,x_{IJ}\Si_{IJ})$.   
In these terms one gets
$$
(X, X)=x_0^2+x_I^2-\frac{1}{2}x_{IJ}^2\,.
$$
There is one more independent invariant combination of $x_0$, $x_I$
and $x_{IJ}$ originating from the invariant $\mbox{tr}(X\bar
X)^2\sim\mbox{det}X$.
Relative to the embedding $SO(5,C)\supset SO(5)$ one has the following
decomposition  in the irreducible representations: 
$$\label{eq:16} 
\un {16}=\un 1\oplus \un 5\oplus\un{10}\,.
$$
Under the discrete transformations  one gets
\bea\label{eq:PT}
P&:&x_0\to x_0,\ x_I\to x_I,\ x_{IJ}\to -x_{IJ}\,,\nn\\
T&:&x_0\to -x_0,\  x_I\to -x_I,\ x_{IJ}\to x_{IJ}\,.\nn
\eea
From the point of view of $SO(5)$, $x_I$ is the
axial vector whereas $x_{IJ}$ is the pseudo-tensor.

The rank of the algebra $C_2$ being $l=2$, an arbitrary irreducible
representation of
the noncompact group $Sp(4,C)$ is uniquely characterized by two
complex Casimir operators $I_2$ and $I_4$ of the second and the forth
order, respectively, i.e.\ by four real quantum numbers. Otherwise, an
irreducible representation of $Sp(4,C)$ can be described by the mixed
spin-tensor $\Psi_{A_1\dots}^{\bar B_1 \dots}$ of a proper rank. This
spin-tensor should be traceless in any pair of the indices of the same
kind, and its symmetry in each kind of indices should correspond
to a two-row Young scheme. In fact,  antisymmetry is
possible in no more than pairs of indices of the same kind. Therefore
an irreducible representation of $Sp(4,C)$  may unambiguously be
characterized by a set of four integers ($r_1,r_2;\bar r_1,\bar r_2$),
$r_1\ge r_2\ge 0$ and $\bar r_1\ge \bar r_2\ge 0$. Here $r_{1,2}$
(respectively, $\bar r_{1,2}$) are the numbers of boxes in the first
or second rows of the proper Young scheme.  The rank of
the maximal compact subgroup $SO(5)\simeq
Sp(4)/Z_2$ (the rotational group) being  equal to $l=2$, a state
in a representation is additionally characterized under fixed boosts
by two additive quantum numbers, namely, the eigenvalues  of the
mutually commuting  momentum components  of $L_{IJ}$ in two different
planes, say, $L_{12}$ and $L_{45}$.

\section{{\boldmath $\Delta$l = 1} reduction}

The ultimate unit of dimensionality in the symplectic
approach is the discrete number $l=1,2,\dots$ corresponding to the
dimensionality $2l$ of the spinor space. The dimensionality
$d=4l^2$ of the \st appears just as a derivative quantity. In reality,
the extended \st with  $l>1$ should supposedly
compactify to the ordinary one with $l=1$ by means of the symplectic
gravity. Let us consider the next-to-ordinary \st case with $l=2$. 
Three generic inequivalent types of the  spinor decomposition relative
to the embedding $Sp(4,C)\supset Sp(2,C)$ are conceivable: 
({\em i})~$\un 4= \un 2\oplus \un 2$,  ({\em ii})~ $\un 4= \un 2\oplus
\ov 2$ and ({\em iii})~$\un 4= \un 2\oplus \un 1\oplus \un 1$.

\vspace{1ex}
({\em i}) Chiral spinor doubling

$$\label{eq:2+2}
\un 4= \un 2\oplus \un 2
$$
results in the decomposition of the Hermitian second-rank spin-tensor
$\un {16}\sim \un 4 \times \ov 4$  as 
$$
\un {16}=4\cdot \un 4\,,
$$
i.e., in a collection of four four-vectors (more precisely, of
three vectors and an axial one). 
As for matter fermions,  the number of the two-component fermions
after compactification is twice that of the number of the
four-component fermions prior compactification. If a
kind of the family structure reproduces itself during the
compactification, it is imperative that there should be at least two
copies of the fermions in the extended space-time, with  at least four
copies of them  in the ordinary space-time.
For phenomenological reasons, the fermions in excess of three families
should acquire rather large effective Yukawa couplings as a
manifestation of curling-up of the space-time extra dimensions. This
is not in principle impossible
because the two-component fermions  distinguish
extra dimensions.  Due to possible appearance of the additional
moderately heavy vector bosons,
the compactification scale $\Lambda$ could in principle be both 
moderate and high without conflict with the standard model
consistency. On the other hand, the  extra time-like dimensions
violate causality and the proper compactification scale~$\Lambda$ in
the pseudo-orthogonal  case should be not less
than the Planck scale~\cite{ynd}. Nevertheless, one may hope that
the latter restriction could be abandoned in the symplectic
approach due to validity of the non-relativistic causality. It is
to be valid at small boosts or gravitational fields, so that the
compactification
scale~$\Lambda$ could possibly be admitted to be not very high. 
For this reason, the given compactification scenario could still
survive at any $\Lambda$.

\vspace{1ex}
({\em ii}) Vector-like spinor doubling

$$\label{eq:2bar2} 
\un 4= \un 2\oplus \ov 2
$$
results in the decomposition
$$
\un {16}=2\cdot \un 4\oplus \Big(\un 3+ \mbox{h.c.}\Big)
\oplus 2\cdot \un 1\,.
$$
In the traditional four-vector notations one has $X\sim
(x_\mu^{(1,2)}$,
$x_{[\mu\nu]}$, $x^{(1,2)}$), $\mu,\nu=0,\dots,3$, with tensor 
$x_{[\mu\nu]}$ being antisymmetric and all the
components $x$ being real. After
compactification there should emerge the  pairs of the
ordinary and mirror matter fermions. For phenomenological reasons, one
should require the mirror fermions to have masses supposedly of the
order of the compactification
scale~$\Lambda$. Modulo reservations for the preceding case, this
compactification scenario could be valid at any~$\Lambda$, too.

\vspace{1ex}
({\em iii}) Spinor-scalar decomposition

$$\label{eq:211}
\un 4= \un 2\oplus \un 1\oplus \un 1
$$
results in 
$$
\un {16}= \un 4\oplus \Big(2\cdot\un 2+  \mbox{h.c.}\Big)
\oplus 4\cdot \un 1\,,
$$
or in the mixed four-vector and spinor notations $X\sim (x_\mu$,
$x_A^{(1,2)}$, $x^{(1,2,3,4)}$), $A=1,2$. There would take place the
violation of the spin-statistics connection  for matter fields in the
four-dimensional space-time if this connection  fulfilled in the
extended space-time. The scale of this violation should be determined
by the compactification scale $\Lambda$ which, in contrast with the
two preceding cases, have safely to be high enough for not to
violate causality within the experimental precision.

\section{Gauge interactions}

Let $D_A{}^{\bar B}\equiv\partial_A{}^{\bar B}+igG_A{}^{\bar B}$ be
the generic covariant derivative, with $g$ being the gauge
coupling, the Hermitian spin-tensor
$G_A{}^{\bar B}$ being  the gauge fields  and $\partial_A{}^{\bar
B}\equiv \partial/
\partial X^A{}_{\bar B}$ being the ordinary derivative. Now let us
introduce the strength tensor
\bea
F_{\{A_1A_2\}}^{[\bar B_1\bar B_2]}&\equiv&
\frac{1}{ig}D_{\{A_1}^{[\bar B_1} D_{A_2\}}^{\bar B_1]}\nn\\
&=&\frac{1}{4ig}\Big(D_{A_1}^{\bar B_1}D_{A_2}^{\bar B_2}-
D_{A_2}^{\bar B_2}D_{A_1}^{\bar B_1}+
D_{A_2}^{\bar B_1}D_{A_1}^{\bar B_2}-
D_{A_1}^{\bar B_2}D_{A_2}^{\bar B_1}\Big)\nn
\eea
and similarly for $\ov F{}_{[A_1A_2]}^{\{\bar B_1\bar B_2\}}\equiv
(F_{\{B_2 B_1\}}^{[\bar A_2\bar A_1]})^*$, where  $\{\dots\}$ and
$[\dots]$ mean the symmetrization and antisymmetrization,
respectively.
One gets
$$
F_{\{A_1A_2\}}^{[\bar B_1\bar B_2]}=
\partial_{\{A_1}^{[\bar B_1} G_{A_2\}}^{\bar B_2]}+
ig G_{\{A_1}^{[\bar B_1} G_{A_2\}}^{\bar B_2]}
$$
and similarly for $\ov F{}_{[A_1A_2]}^{\{\bar B_1\bar B_2\}}$. These
tensors are gauge invariant. The total number of the real
components in  the tensor
$F_{\{A_1A_2\}}^{[\bar B_1\bar B_2]}$ precisely coincides with the
number of components of the antisymmetric second-rank tensor
$F_{[\alpha\beta]}$, $\alpha,\beta=0,1,\dots,4l^2-1$ defined in the
pseudo-Euclidean space of the  $d=4l^2$ dimensions. But in the
symplectic case, tensor $F$ is
reducible and splits into a trace relative to $\e$ and a traceless
part, $F=F^{(0)}+F^{(1)}$, where  
$F^{(0)}{}_{\{A_1A_2\}}^{[\bar B_1\bar B_2]}\equiv
F^{(0)}_{\{A_1A_2\}}\e^{\bar B_1\bar B_2}$ and  
$F^{(1)}{}_{\{A_1A_2\}}^{[\bar B_1\bar B_2]}\e_{\bar B_1\bar B_2}=0$
(and similarly for $\ov F_{[A_1A_2]}^{\{\bar B_1\bar B_2\}}$). 

For an unbroken gauge theory with fermions, the generic gauge, fermion
and mass terms of the Lagrangian ${\cal L}= {\cal L}_G+{\cal
L}_F+{\cal L}_M$ are, respectively,
\bea\label{eq:L}
{\cal L}_G&=&\sum_{s=0,1}(c_s +i\theta_s)\,
F^{(s)}F^{(s)}+{\rm h.c.} \,,\nn\\[-0.5ex]
{\cal L}_F&=&\frac{i}{2}\sum_{\pm}(\psi ^{\pm})^\dagger\!
\stackrel{\leftrightarrow} D \psi^{\pm}\,,\nn\\
{\cal L}_M&=&\psi^+m_0\,\psi^-
+\sum_{\pm}\psi^{\pm}m_{\pm}\psi^{\pm}+{\rm h.c.}\,,\nn
\eea
where $F^{(s)}F^{(s)}\equiv 
F^{(s)}{}_{\{A_1A_2\}}^{[\bar B_1\bar B_2]}
F^{(s)}{}^{\{A_2A_1\}}_{[\bar B_2\bar B_1]}$.
In the Lagrangians above, $\psi^{\pm}$ are the
charged conjugate fermions, $m_0$ is the generic Dirac mass, $m_\pm$
are Majorana masses, $c_s$ and $\theta_s$ are the  real gauge
parameters. One of the parameters $c_s$, supposedly
$c_0\neq 0$, can be normalized at will. 
The Lagrangian results in the following generalization of the Dirac
equation
$$\label{eq:dir}
iD^C\!{}_{\bar B}\psi^{\pm}_C= m_0^\dagger \ov\psi{}^{\pm}_{\bar B}
+\sum_{\pm} m_\pm^\dagger\, \ov \psi{}^{\mp}_{\bar B}
$$
and the pair of Maxwell equations ($c_0\equiv 1$ and $c_1=\theta_1=0$,
for simplicity)
\bea\label{eq:max}
(1 +i\theta_0)D^{C\bar B}F^{(0)}{}_{\{C A
\}}-\mbox{h.c.}&=&0\,,\nn\\
(1 +i\theta_0)D^{C\bar B}F^{(0)}{}_{\{C A\}}+\mbox{h.c.}
&=&2 g J_{A}{}^{\bar B}\,,\nn
\eea
with 
$J_{A}{}^{\bar B}\equiv\sum_{\pm} (\pm 1)
\psi^{\pm}_{A}({\psi}{}^\pm_{B})^\dagger$
being the fermion Hermitian current.

Tensors $F^{(s)}$, $s=1,2$ are non-Hermitian, but 
under restriction by the maximal compact subgroup $Sp(2l)$ 
they  split into a pair of the Hermitian ones $E^{(s)}$ and $B^{(s)}$
as $F^{(s)}=E^{(s)}+iB^{(s)}$. 
Introducing the duality transformation
$F^{(s)}\to \tilde F^{(s)}\equiv -iF^{(s)}$, so that $\tilde
E^{(s)}=B^{(s)}$ and
$\tilde B^{(s)}=-E^{(s)}$, one gets 
${\cal R}e F^{(s)}F^{(s)}=E^{(s)}{}^2-B^{(s)}{}^2$ and 
${\cal I}m F^{(s)}F^{(s)}={\cal R}e \tilde
F^{(s)}F^{(s)}=2E^{(s)}B^{(s)}$. 
Though the splitting into $E^{(s)}$ and $B^{(s)}$ is noncovariant
with respect to the whole $Sp(2l,C)$, the duality transformation is
covariant. Tensors $E^{(s)}$ and $B^{(s)}$ are the counterparts of
the ordinary electric and magnetic strengths,  and $\theta_0$ is the
counterpart of the ordinary $T$-violating $\theta$-parameter for the
$l=1$ case. Thus, $\theta_1$ is an additional $T$-violating parameter
at $l>1$. Note that in the framework of  symplectic extension the
electric and magnetic strengths stay on equal footing. This is to be
contrasted with the pseudo-orthogonal extension where these strengths
have  unequal number of components at $d\neq 4$.
The electric-magnetic duality for the
gauge fields (in the Euclidean space) play an important role  for the
study of the topological structure of the  gauge vacuum in 
four space-time dimensions. Therefore the similar study might be
applicable to the extended symplectic space-times with
arbitrary $l>1$. 

\section{Gravity}

The field equations above are valid in the flat extended \st or,
otherwise, refer to the inertial local frames. To go beyond, 
one can introduce  the  local 
fielbeins $e_M{}_A{}^{\bar B}(X)$, such that $e_M{}_A{}^{\bar
B}=(e_M{}_B{}^{\bar A})^*$,  the real world coordinates $x_M\equiv
e_M{}^A{}_{\bar B} X_A{}^{\bar B}$, as well as the
generally covariant derivative $\nabla\!_M(e)$, with $M=0,1,\dots,
4l^2-1$ being the world vector index. Now, the
Lagrangian can be adapted to the $d=4l^2$ dimensional curved  \st
equipped with the pseudo-Riemannian structure, i.e., the real
symmetric metrics $g_{MN}(x)= e_M{}^A\!{}_{\bar B}
e_N{}_{A}{}^{\bar B}$. One can also supplement gauge equations by the
generalized gravity equations in line with~\cite{uti}.
But in the case at hand, the group of equivalence of the local
fielbeins (structure group) is required to be only the symplectic
group $Sp(2l,C)$ rather than the whole pseudo-orthogonal group
$SO(d_-,d_+)$. The former one leaves more independent components in
the local symplectic fielbeins compared to  the  pseudo-Riemannian
fielbeins and thus to the metrics. Hence the symplectic gravity is not
in general equivalent to the metric one. 
The curvature tensor in the symplectic case, like the
gauge strength one, splits additionally into irreducible parts which
can a~priori enter the gravity Lagrangian with the independent
coefficients. The ultimate reason for this 
is that  the \st is meant to be in the symplectic approach
not a fundamental entity. Therefore gravity as a generally
covariant theory of the \st deformations have to be just as an
effective theory. The latter admits the existence of a  number of free
parameters, the choice of which should eventually be
clarified by an underlying theory.

\section{Conclusion}

The hypothesis that the symplectic structure of \st is superior to the
metric one provides, in particular, the rationale for the
four-dimensionality and the $1+3$ signature of the ordinary
space-time. When looking for the space-times with extra dimensions,
the hypothesis  predicts the discrete series of the metric space-times
of the peculiar dimensionalities and signatures, both with the 
spatial and  time extra dimensions. One of the time-like directions
remaining rotationally invariant under fixed boosts makes it possible
the approximate causality description despite the presence
of extra times. The extended symplectic space-times provide a viable
alternative to the pseudo-orthogonal ones. But beyond the physical
adequacy of the extended space-times as such, by gene\-ralizing
from
the
basic case $l=1$ to its counterpart for general $l>1$, a deeper
insight into the nature of the four-dimensional \st itself may be
gained.

\paragraph{Acknowledgement}
The author is grateful to the  Organizing Committee for support.

\end{document}